\documentclass[12pt]{iopart}

\usepackage{iopams}

\usepackage{hyperref}
\usepackage{tabularx}
\usepackage{soul}
\usepackage{xcolor}
\usepackage{adjustbox}
\usepackage{cite}
\begin{document}

\title[Cosmological Parameter Estimation]{On Direct Estimation of Density Parameters and Hubble Constant for $\Lambda$CDM Universe using Hubble Measurements}

%\author{Srikanta Pal$^{1,a}$\orcidlink{0000-0002-9502-8510} and Rajib Saha$^{1,b}$\orcidlink{0000-0002-4444-1081}}
\author{Srikanta Pal$^{1,a}$ and Rajib Saha$^{1,b}$}

\address{$^1$Department of Physics, Indian Institute of Science Education and Research Bhopal, Bhopal-462066, Madhya Pradesh, India}
\eads{\mailto{$^a$srikanta18@iiserb.ac.in}, \mailto{psrikanta357@gmail.com}}
%\ead{$^a$srikanta18@iiserb.ac.in}
\ead{$^b$rajib@iiserb.ac.in}
\vspace{10pt}
\begin{indented}
\item[]June 2024
\end{indented}

\begin{abstract}
The set of cosmological density parameters ($\Omega_{0m}h_{0}^{2}$, $\Omega_{0k}h_{0}^{2}$, $\Omega_{0\Lambda}h_{0}^{2}$) and Hubble constant ($\hat{h}_{0}$) are useful for fundamental understanding of the universe from many perspectives. In this article, we propose a new procedure to estimate these parameters for $\Lambda$ cold dark matter ($\Lambda$CDM) universe in the Friedmann-Robertson-Walker (FRW) background. We generalize the two-point statistics first proposed by Sahni \etal (2008) to the three point case and estimate the parameters using currently available Hubble parameter ($H(z)$) values in the redshift range $0.07 \leq z \leq 2.36$ measured by \textit{differential age} (DA) and \textit{baryon acoustic oscillation} (BAO) techniques. All the parameters are estimated assuming the general case of non-flat universe. Using  both DA and BAO data we obtain $\Omega_{0m}h_{0}^{2}=0.1485 \pm 0.0065$, $\Omega_{0k}h_{0}^{2}=-0.0137 \pm 0.017$, $\Omega_{0\Lambda}h_{0}^{2}=0.3126 \pm 0.0145$ and $\hat{h}_{0}=0.6689 \pm 0.0021$. These results are in satisfactory agreement with the Planck results. An important advantage of our method is that to estimate the value of any one of the independent cosmological parameters one does not need to use the values for the rest of them. Each parameter is obtained solely from the measured values of Hubble parameters at different redshifts without any need to use values of other parameters. Such a method is expected to be less susceptible to the undesired effects of degeneracy issues between cosmological parameters during their estimations. Moreover, there is no requirement of assuming spatial flatness in our method.
\end{abstract}
%\keywords{Cosmological observations, cosmological parameters, statistical methods, numerical analysis}
\vspace{2pc}
\noindent{\it Keywords}: Cosmological observations, cosmological parameters, statistical methods, numerical analysis
%\maketitle

\section{Introduction}
The cosmological principle states that our universe is {\it homogeneous} and {\it isotropic} on a large enough scale ($\gtrsim 300$ Mpc). The recent observations of {\it cosmic microwave background} (CMB) radiation and its interpretations establish the explanations for the accelerated expansion, dark energy domination and flatness of the universe~\cite{Planck_2018}. the $\Lambda$CDM model is known as the standard model of Big Bang cosmology. This is the simplest model of our universe considering three fundamental density components, such as cosmological constant ($\Lambda$), cold dark matter and visible matter density.

Dark matter and dark energy are two mysterious components of the universe. Although we are familiar with some properties of these components, an understanding at a fundamental level about them is lacking till date. Dark matter has zero pressure, same as ordinary matter, and it interacts with nothing except gravitation. Dark energy, which perhaps causes the accelerated expansion of the universe~\cite{Riess_1998,Perlmutter_1999}, has negative pressure. the cosmological constant ($\Lambda$), first envisioned by Albert Einstein almost a century ago~\cite{Einstein_1917}.
%is treated as the simplest form of dark energy which has the pressure exactly equal to the density with a negative sign.
In order not to represent a medium absolutely at rest, violating the principle of relativity, it was in 1933 interpreted by Lemaitre as representing the constant density of Lorentz Invariant Vacuum Energy (LIVE)~\cite{Lemaitre_1933,Lemaitre_1997,Zeldovich_1968,Gron_1986}. It follows from the requirement that all the components of the energy-momentum tensor shall be Lorentz invariant, that the pressure of LIVE is equal to minus its energy density, and hence that LIVE causes repulsive gravity. Recent observations show that our universe contains $4.6\%$ visible baryonic matter, $24\%$ dark matter and $71.4\%$ LIVE approximately\footnote{\url{https://map.gsfc.nasa.gov/media/121236/index.html}}.

CMB data shows an excellent agreement with $\Lambda$CDM model of the universe. However, many research projects show some disagreement to accept the $\Lambda$CDM model as a final interpretable model of the universe. Zunckel \& Clarkson (2008)~\cite{Zunckel_2008} formulated a `litmus test' to verify the acceptance of $\Lambda$CDM model. They showed a significant deviation of the equation of state of dark energy for $\Lambda$CDM model as well as other dark energy models in their analysis. In the literatures by Macaulay \etal(2013)~\cite{Macaulay_2013}  and Raveri (2016)~\cite{Raveri_2016}, Canada-France-Hawaii-Telescope Lensing Survey measurements also showed a tension with the results of Planck Collaboration VI (2020)~\cite{Planck_2018}. Testing of the Copernican principle~\cite{Uzan_2008,Valkenburg_2014} allows us to understand the existence as well as the evolution of dark energy. These types of concerns encourage the researchers to test the reliability of $\Lambda$CDM model.

Two interesting powerful procedures are $Om$ and $Omh^{2}$ diagnostics for a null test of the $\Lambda$CDM model. Sahni \etal(2008)~\cite{Sahni_2008} developed $Om(z)$ diagnostic to express the density components of the universe in terms of redshift and corresponding Hubble parameter for flat $\Lambda$CDM universe. $Om(z)$ is the matter density parameter ($\Omega_{0m}$) expressed as a combination of Hubble parameter and cosmological redshift. Sahni \etal(2008)~\cite{Sahni_2008} also defined two-point diagnostic which is given by
\begin{eqnarray}
Om(z_{i},z_{j}) &=& Om(z_{i})-Om(z_{j}) \ . \label{Om_two_point}
\end{eqnarray}
They used this two-point diagnostic for null test of flat $\Lambda$CDM model of the universe. Shafieloo \etal(2012)~\cite{Shafieloo_2012} modified the two-point diagnostic expressed in Eqn.~\ref{Om_two_point}. This modified two-point diagnostic is given by
\begin{eqnarray}
Omh^{2}(z_{i},z_{j}) &=& \frac{h^2(z_i)-h^2(z_j)}{(1+z_i)^3-(1+z_j)^3} \ , \label{Omh2_two_point}
\end{eqnarray}
where $Omh^{2}$ is the estimator of $\Omega_{0m}h_{0}^{2}$. Furthermore, $h_{0}$ is the Hubble constant in $100 \ \rm{kmMpc^{-1}sec^{-1}}$ unit. Following Eqn.~\ref{Omh2_two_point}, $Omh^{2}(z_{i},z_{j})$ is a powerful probe for null test of the flat $\Lambda$CDM universe~\cite{Shafieloo_2012,Sahni_2014}. Sahni \etal(2014)~\cite{Sahni_2014} utilized the two-point diagnostic $Omh^{2}(z_{i},z_{j})$ for the null test using three Hubble parameter measurements ($H(z)$) estimated by {\it baryon acoustic oscillation} (BAO) technique. These specific three $H(z)$ are $H(z=0)$ measured by Riess \etal(2011)~\cite{Riess_2011} and the Planck Collaboration XVI (2014)~\cite{Planck_2013}, $H(z=0.57)$ from Sloan Digital Sky Survey Data Release $9$ (SDSS DR$9$)~\cite{Samushia_2013}, and $H(z=2.34)$ from Ly$\alpha$ forest in SDSS DR$11$~\cite{Delubac_2015}. Using these three $H(z)$, Sahni \etal(2014)~\cite{Sahni_2014} showed that their null test experiment gives a strong tension with the value of $\Omega_{0m}h_{0}^{2}$ from the Planck Collaboration XVI (2014)~\cite{Planck_2013}. $Om(z_{i}, z_{j})$ and $Omh^{2}(z_{i}, z_{j})$ diagnostics are also applied by Zheng \etal(2016)~\cite{Zheng_2016} for three different models ($\Lambda$CDM, wCDM and Chevalier-Polarski-Linder(CPL;~\cite{Chevalier_2001,Linder_2003})) of the flat universe. They also found significant tension with the Planck Collaboration XVI (2014)~\cite{Planck_2013} results, for each of these three models.

In the current article, we ask the following question. \textit{Is it possible to measure the density parameters separately and yet uniquely using a set of observed Hubble parameter values at different redshifts?} The method estimates each density parameter directly from the data using some unique mapping functions\footnote{The mapping functions are described in detail in section~\ref{three_point}.} which of course are different for different density parameters. \textit{We use} $\Lambda$CDM \textit{universe but we do not assume any prior information about the curvature, keeping the flexibility to probe the density parameters individually in a general constant spatial curvature model.} In our analysis, we utilize $31$ Hubble parameters measured by DA technique in the redshift range $0.07 \leq z \leq 1.965$ as well as $24$ Hubble parameters measured by BAO technique in the redshift range $0.24 \leq z \leq 2.36$. We note that DA technique does not assume any cosmological model to estimate the Hubble parameter. Additionally, BAO technique assumes flat $\Lambda$CDM model to estimate the sound horizon which is used to obtain the corresponding Hubble parameter. We describe these observed Hubble data in section~\ref{data}. Using combinations of three measured Hubble parameter values at a time, we estimate several values for matter density ($\Omega_{0m}h_{0}^{2}$), curvature density ($\Omega_{0k}h_{0}^{2}$) and cosmological constant density ($\Omega_{0\Lambda}h_{0}^{2}$). We call the statistics employed for such estimations as `three point statistics', their values as `sample specific values' and the uncertainties corresponding to their values as `sample specific uncertainties' in this article. \textit{An important advantage of our method is that to estimate values of any cosmological density parameter one does not have to assume values for the others. Thus our results are data-driven to a significant extent. We note that, our method can be seen as a generalization of the two point statistics proposed earlier in the literatures by Shafieloo \etal(2012)}~\cite{Shafieloo_2012} \textit{and Sahni \etal(2014)}~\cite{Sahni_2014}. \textit{Thus, our method is completely parameter independent.} We utilize median statistics in our analysis to estimate the median value of each density parameter and the corresponding uncertainty from the smaple specific values of these density parameters. After finding the median values with the corresponding uncertainties for the three density parameters (along with their covariances), we estimate the value of Hubble constant. We notice that the uncertainties in the sample specific values of each density parameter are non-Gaussian in nature. Therefore, we do not use weighted-mean statistics in our analysis, since this statistics uses the uncertainties of measurements for the estimation of mean assuming the Gaussian nature of these uncertainties~\cite{Zheng_2016}. We note that median statistics is appropriate for our analysis because the uncertainties of measurements are not used in this statistics~\cite{Gott_III_2001,Zheng_2016}. Therefore, the non-Gaussianity of sample specific uncertainties does not affect the parameter values estimated by using median statistics. Using median statistics and using all (DA+BAO) $53$ Hubble measurements, we find that our best estimates of the cosmological parameters are in excellent agreement with the Planck results~\cite{Planck_2018}.

Our paper is arranged as follows. In section~\ref{three_point} we describe the basic equations for all the cosmological parameters estimated by us using measured Hubble parameter values in different redshifts. We discuss about median statistics in section~\ref{md_stat}. In section~\ref{data}, we show the Hubble parameter measurements used in our analysis. In sections~\ref{da_tech} and~\ref{bao_tech}, we give a brief overview about DA and BAO techniques. In section~\ref{num_val}, we discuss the three-point statistics for processing of sample specific values of density parameters. We discuss the non-Gaussian nature of uncertainties corresponding to sample specific values of density parameters in section~\ref{non_gauss}. In section~\ref{mean_val}, we present our estimated results and corresponding uncertainty ranges for cosmological parameters. In section~\ref{mean_val}, we show the Hubble parameter curves using our estimated results as well as the results of Planck Collaboration VI (2020)~\cite{Planck_2018}. Finally, in section~\ref{conclusions}, we conclude our analysis.

\section{Formalism} 
\label{formalism}
\subsection{Three-point statistics}
\label{three_point}
The well known Einstein's field equation is given by
\begin{eqnarray}
R_{\mu \nu}-\frac{1}{2}g_{\mu \nu}R = -\frac{8 \pi G}{c^{4}}T_{\mu \nu} \ , \label{field_eqn}
\end{eqnarray}
where $R_{\mu \nu}$ is Ricci curvature tensor, $R$ is Ricci curvature scalar, $g_{\mu \nu}$ is metric tensor, $T_{\mu \nu}$ is energy-momentum tensor, $G$ is universal gravitational constant and $c$ is the velocity of light in vacuum.
 
The Friedmann-Robertson-Walker (FRW) line element, in spherical coordinate system, can be expressed as
\begin{eqnarray}
ds^{2} = c^{2}dt^{2}-a^{2}(t)\left[\frac{dr^{2}}{1-kr^{2}}+r^{2}\left(d\theta ^{2}+\sin^{2}\theta d\phi^{2}\right)\right], \label{FRW}
\end{eqnarray}
where $r , \theta , \phi$ are comoving co-ordinates. the scale factor of the universe is defined by $a(t)$ and the curvature constant is denoted by $k$. Zero value of $k$ defines the spatially flat universe. Positive value of $k$ represents that the universe is closed and negative value of $k$ signifies that the universe is open. 

Using Eqns.~\ref{field_eqn} and~\ref{FRW}, we find two Friedmann equations which can be written as
\begin{eqnarray}
\frac{\dot{a}^{2}}{a^{2}}+\frac{kc^{2}}{a^{2}} = \frac{8 \pi G}{3}\rho(t), \label{friedmann_eqn1} 
\\
\frac{2\ddot{a}}{a}+\frac{\dot{a}^{2}}{a^{2}}+\frac{kc^{2}}{a^{2}} = -\frac{8 \pi G}{c^{2}}P(t), \label{friedmann_eqn2}
\end{eqnarray}
where $\rho(t)$ is the density and $P(t)$ is the gravitational pressure of the universe. In Eqns.~\ref{friedmann_eqn1} and~\ref{friedmann_eqn2}, $\dot{a}$ represents the first order time derivative of scale factor and $\ddot{a}$ defines the second order time derivative of scale factor.

From these two Friedmann equations (Eqns.~\ref{friedmann_eqn1} and~\ref{friedmann_eqn2}) for $\Lambda$CDM model (neglecting radiation density term for late time universe), the Hubble parameter as a function of redshift is defined by
\begin{eqnarray}
H^{2}(z) = H^{2}_{0}\left[\Omega _{0m}(1+z)^{3}+\Omega _{0k}(1+z)^{2}+\Omega_{0\Lambda}\right]. \label{H_z}
\end{eqnarray}
The $density$ $parameter$ ($\Omega$) is specified by the ratio between density ($\rho$) and critical density ($\rho_{c}$). The critical density ($\rho_{c}$) is expressed as $3H^{2}/8\pi G$. In Eqn.~\ref{H_z}, $\Omega _{0m}$ is the matter density parameter, $\Omega _{0k}$ (defined as $-kc^{2}/a_0^{2}H_0^{2}$) is the curvature density parameter and $\Omega _{0\Lambda}$ is the cosmological constant density parameter. We can also explain the spatial curvature of the universe by looking into the value of $\Omega _{0k}$. If $\Omega _{0k}=0$, then universe is spatially flat. Positive and negative values of $\Omega _{0k}$ indicate that universe is open and closed respectively.

Defining $h(z) = \frac{H(z)}{100} \ 100 \rm{kmMpc^{-1}sec^{-1}}$ and $h_{0} = \frac{H_{0}}{100} \ 100 \rm{kmMpc^{-1}sec^{-1}}$, the Hubble parameter expression at $\alpha$-th redshift can be written as
\begin{eqnarray}
h^{2}(z_{\alpha}) &=& \Omega _{0m}h_{0}^{2}(1+z_{\alpha})^{3}+\Omega _{0k}h_{0}^{2}(1+z_{\alpha})^{2}+\Omega_{0\Lambda}h_{0}^{2}. \label{h_z_i}
\end{eqnarray}
Similarly, the Hubble parameter expressions, at $\beta$-th and $\gamma$-th redshifts, are given by
\begin{eqnarray}
h^{2}(z_{\beta}) &=& \Omega _{0m}h_{0}^{2}(1+z_{\beta})^{3}+\Omega _{0k}h_{0}^{2}(1+z_{\beta})^{2}+\Omega_{0\Lambda}h_{0}^{2},\label{h_z_j} \\
h^{2}(z_{\gamma}) &=& \Omega _{0m}h_{0}^{2}(1+z_{\gamma})^{3}+\Omega _{0k}h_{0}^{2}(1+z_{\gamma})^{2}+\Omega_{0\Lambda}h_{0}^{2}.\label{h_z_k}
\end{eqnarray} 
Inevitably, we have three equations (Eqns.~\ref{h_z_i},~\ref{h_z_j} and~\ref{h_z_k}) now. Three unkonwn coefficients corresponding to these equations are $\Omega_{0m}h_{0}^{2}$, $\Omega_{0k}h_{0}^{2}$ and $\Omega_{0\Lambda}h_{0}^{2}$. These unknown coefficients can be easily resolved from these three equations.

The matter density parameter ($\Omega_{0m}h_{0}^{2}$), using Eqns.~\ref{h_z_i},~\ref{h_z_j} and~\ref{h_z_k}, can be expressed by
\begin{eqnarray}
\Omega_{0m}h_{0}^{2} &=& \frac{h^{2}(z_{\alpha})\left[(1+z_{\beta})^{2}-(1+z_{\gamma})^{2}\right]}{-A(z_{\alpha},z_{\beta},z_{\gamma})}+\frac{h^{2}(z_{\beta})\left[(1+z_{\gamma})^{2}-(1+z_{\alpha})^{2}\right]}{-A(z_{\alpha},z_{\beta},z_{\gamma})}\nonumber \\ & & +\frac{h^{2}(z_{\gamma})\left[(1+z_{\alpha})^{2}-(1+z_{\beta})^{2}\right]}{-A(z_{\alpha},z_{\beta},z_{\gamma})}. \label{Omh2}
\end{eqnarray}
The curvature density parameter ($\Omega_{0k}h_{0}^{2}$), using Eqns.~\ref{h_z_i},~\ref{h_z_j} and~\ref{h_z_k}, is given by
\begin{eqnarray}
\Omega_{0k}h_{0}^{2} & = & \frac{h^{2}(z_{\alpha})\left[(1+z_{\beta})^{3}-(1+z_{\gamma})^{3}\right]}{A(z_{\alpha},z_{\beta},z_{\gamma})}+\frac{h^{2}(z_{\beta})\left[(1+z_{\gamma})^{3}-(1+z_{\alpha})^{3}\right]}{A(z_{\alpha},z_{\beta},z_{\gamma})}\nonumber \\ & & +\frac{h^{2}(z_{\gamma})\left[(1+z_{\alpha})^{3}-(1+z_{\beta})^{3}\right]}{A(z_{\alpha},z_{\beta},z_{\gamma})}. \label{Okh2}
\end{eqnarray}
The cosmological constant density parameter ($\Omega_{0\Lambda}h_{0}^{2}$), using Eqns.~\ref{h_z_i},~\ref{h_z_j} and~\ref{h_z_k}, can be written as
\begin{eqnarray}
\hspace{-40pt}
\Omega_{0\Lambda}h_{0}^{2} &=& \frac{h^{2}(z_{\alpha})(1+z_{\beta})^{2}(1+z_{\gamma})^{2}(z_{\gamma}-z_{\beta})}{A(z_{\alpha},z_{\beta},z_{\gamma})}+\frac{h^{2}(z_{\beta})(1+z_{\gamma})^{2}(1+z_{\alpha})^{2}(z_{\alpha}-z_{\gamma})}{A(z_{\alpha},z_{\beta},z_{\gamma})}\nonumber \\ &&+\frac{h^{2}(z_{\gamma})(1+z_{\alpha})^{3}(1+z_{\beta})^{3}(z_{\beta}-z_{\alpha})}{A(z_{\alpha},z_{\beta},z_{\gamma})}, \label{Odh2}
\end{eqnarray}
where
\begin{eqnarray}
\hspace{-40pt}
A(z_{\alpha},z_{\beta},z_{\gamma}) &=& (1+z_{\alpha})^{2}(1+z_{\beta})^{2}(z_{\beta}-z_{\alpha})+(1+z_{\beta})^{2}(1+z_{\gamma})^{2}(z_{\gamma}-z_{\beta})\nonumber \\ && +(1+z_{\gamma})^{2}(1+z_{\alpha})^{2}(z_{\alpha}-z_{\gamma}). \label{A_z1z2z3}
\end{eqnarray} 

Assuming independent measurement of Hubble parameter at each redshift and using error propagation formula in Eqns.~\ref{Omh2},~\ref{Okh2} and~\ref{Odh2}, the uncertainties corresponding to $\Omega_{0m}h_{0}^{2}$, $\Omega_{0k}h_{0}^{2}$ and $\Omega_{0\Lambda}h_{0}^{2}$ are given by
\begin{eqnarray}
\sigma_{\Omega_{0m}h_{0}^{2}}^{2} & = &\frac{4h^{2}(z_{\alpha})\sigma^{2}_{h(z_{\alpha})}\left[(1+z_{\beta})^{2}-(1+z_{\gamma})^{2}\right]^{2}}{A^{2}(z_{\alpha},z_{\beta},z_{\gamma})}\nonumber \\ && +\frac{4h^{2}(z_{\beta})\sigma^{2}_{h(z_{\beta})}\left[(1+z_{\gamma})^{2}-(1+z_{\alpha})^{2}\right]^{2}}{A^{2}(z_{\alpha},z_{\beta},z_{\gamma})}\nonumber \\ && +\frac{4h^{2}(z_{\gamma})\sigma^{2}_{h(z_{\gamma})}\left[(1+z_{\alpha})^{2}-(1+z_{\beta})^{2}\right]^{2}}{A^{2}(z_{\alpha},z_{\beta},z_{\gamma})},\label{sigma_Omh2}
\end{eqnarray}
\begin{eqnarray}
\sigma_{\Omega_{0k}h_{0}^{2}}^{2} & = &\frac{4h^{2}(z_{\alpha})\sigma^{2}_{h(z_{\alpha})}\left[(1+z_{\beta})^{3}-(1+z_{\gamma})^{3}\right]^{2}}{A^{2}(z_{\alpha},z_{\beta},z_{\gamma})}\nonumber \\ & & +\frac{4h^{2}(z_{\beta})\sigma^{2}_{h(z_{\beta})}\left[(1+z_{\gamma})^{3}-(1+z_{\alpha})^{3}\right]^{2}}{A^{2}(z_{\alpha},z_{\beta},z_{\gamma})}\nonumber \\ & & +\frac{4h^{2}(z_{\gamma})\sigma^{2}_{h(z_{\gamma})}\left[(1+z_{\alpha})^{3}-(1+z_{\beta})^{3}\right]^{2}}{A^{2}(z_{\alpha},z_{\beta},z_{\gamma})},\label{sigma_Okh2}
\end{eqnarray}
\begin{eqnarray}
\sigma_{\Omega_{0\Lambda}h_{0}^{2}}^{2} &=& \frac{4h^{2}(z_{\alpha})\sigma^{2}_{h(z_{\alpha})}(1+z_{\beta})^{4}(1+z_{\gamma})^{4}(z_{\gamma}-z_{\beta})^{2}}{A^{2}(z_{\alpha},z_{\beta},z_{\gamma})}\nonumber \\ && +\frac{4h^{2}(z_{\beta})\sigma^{2}_{h(z_{\beta})}(1+z_{\gamma})^{4}(1+z_{\alpha})^{4}(z_{\alpha}-z_{\gamma})^{2}}{A^{2}(z_{\alpha},z_{\beta},z_{\gamma})}\nonumber \\ && +\frac{4h^{2}(z_{\gamma})\sigma^{2}_{h(z_{\gamma})}(1+z_{\alpha})^{4}(1+z_{\beta})^{4}(z_{\beta}-z_{\alpha})^{2}}{A^{2}(z_{\alpha},z_{\beta},z_{\gamma})}. \label{sigma_Odh2}
\end{eqnarray}

The Hubble constant ($\hat{h}_{0}$), Hubble parameter at redshift $z=0$, can be expressed by
\begin{eqnarray}
\hat{h}_{0}^{2} & = & \Omega _{0m}h_{0}^{2}+\Omega _{0k}h_{0}^{2}+\Omega_{0\Lambda}h_{0}^{2}. \label{h_z_0}
\end{eqnarray}

The uncertainty corresponding to the Hubble constant ($\hat{h}_{0}$), using error propagation formula in Eqn.~\ref{h_z_0}, is given by
\begin{eqnarray}
\sigma_{\hat{h}_{0}}^{2} &=& \frac{1}{4h_{0}^{2}}\biggl[\sigma_{\Omega_{0m}h_{0}^{2}}^{2}+\sigma_{\Omega_{0k}h_{0}^{2}}^{2}+\sigma_{\Omega_{0\Lambda}h_{0}^{2}}^{2}+2\biggl\{{\rm Cov}\bigl(\Omega_{0m}h_{0}^{2},\Omega_{0k}h_{0}^{2}\bigr)\nonumber \\ && \hspace{20pt}+{\rm Cov}\bigl(\Omega_{0m}h_{0}^{2},\Omega_{0\Lambda}h_{0}^{2}\bigr)+{\rm Cov}\bigl(\Omega_{0k}h_{0}^{2},\Omega_{0\Lambda}h_{0}^{2}\bigr)\biggr\}\biggr], \label{sigma_h0}
\end{eqnarray}
where $\sigma_{\Omega_{0m}h_{0}^{2}}, \ \sigma_{\Omega_{0m}h_{0}^{2}}, \ \sigma_{\Omega_{0m}h_{0}^{2}}$ are $1\sigma$ uncertainties corresponding to the best estimates of the cosmological density parameters. In Eqn.~\ref{sigma_h0}, ${\rm Cov}(x,y)$ denotes the covariance between the samples $x$ and $y$. In our analysis, $x$ and $y$ are the density parameters. So we can easily calculate the value of $\hat{h}_{0}$ and $\sigma_{\hat{h}_{0}}$, from Eqns.~\ref{h_z_0} and~\ref{sigma_h0}, using our best estimates of $\Omega_{0m}h_{0}^{2}$, $\Omega_{0k}h_{0}^{2}$ and $\Omega_{0\Lambda}h_{0}^{2}$.

For $n$ numbers of samples of redshift and corresponding $H(z)$ measurement, we can generate $^{n}C_{3}=n(n-1)(n-2)/6$ numbers of sample specific values for each of the coefficients $\Omega_{0m}h_{0}^{2}$, $\Omega_{0k}h_{0}^{2}$ and $\Omega_{0\Lambda}h_{0}^{2}$~\cite{Zheng_2016}. We analyse these sample specific values for each of three density parameters using median statistics. Theoretically, we should obtain the same values in every calculation for each of the coefficients. Since we use the observed $H(z)$ data in our analysis, we can't obtain the same results in each calculation of numerical analysis.

\subsection{Median Statistics}
\label{md_stat}
Median statistics~\cite{Gott_III_2001, Zheng_2016} is an excellent approach for analysing a large number of samples without assuming the Gaussian nature of the uncertainties corresponding to samples. In this statistics, we consider that all data points are statistically independent and they have no systematic errors. Let us assume that we have total $(N+1)$ number of measurements $\{X_{i} : i \in \{0,1,...,N\}\}$ arranged in ascending order. If $N$ is even, the {\it median} value will be the $\frac{N}{2}$-th measurement. If $N$ is odd then the {\it median} value will be the simple average of $\frac{N-1}{2}$-th and $\frac{N+1}{2}$-th measurements. Let us assume the median of $\{X_i\}$ is $X_{\rm md}$. Then, the variance corresponding to this median~\cite{Muller_2000} can be expressed as
\begin{eqnarray}
\sigma_{X_{\rm md}}^{2} &=& {\rm C^2} \left[{\rm MAD}(X_{\rm md})\right]^2, \label{md_var}
\end{eqnarray}
where ${\rm C}=1.9/\sqrt{N}$ and ${\rm MAD}(X_{\rm md})$ denotes ``median of the absolute deviations'' which can be written as
\begin{eqnarray}
{\rm MAD}(X_{\rm md}) &=& {\rm median \ of \ } \{|X_i-X_{\rm md}|\}. \label{mad}
\end{eqnarray}
Moreover, let us assume another sample $\{Y_i\}$ with median $Y_{\rm md}$, where $\{Y_i\}$ has the same sample size as $\{X_i\}$. Müller (2005)~\cite{Muller_2005} has shown that the covariance between the median values of these two samples can be expressed as
\begin{eqnarray}
{\rm Cov}(X_{\rm md}, Y_{\rm md}) &=& {\rm C}^2\left[{\rm MAC}(X_{\rm md}, Y_{\rm md})\right], \label{md_cov}
\end{eqnarray}
where ${\rm MAC}(X_{\rm md}, Y_{\rm md})$ defines the ``median analogue to covariance'' which can be written as
\begin{eqnarray}
{\rm MAC}(X_{\rm md}, Y_{\rm md}) &=& {\rm median \ of \ } \{(X_i-X_{\rm md})(Y_i-Y_{\rm md})\}. \label{mac}
\end{eqnarray}
We note that the self-covariance estimated by using Eqn.~\ref{md_cov} will be exactly same with the result obtained from Eqn.~\ref{md_var}. We use Eqn.~\ref{md_var} to estimate the variances corresponding to the median values of density parameters. Similarly, we use Eqn.~\ref{md_cov} to obtain the covariances between the median values of density parameters. We utilize these variances and covariances in Eqn.~\ref{sigma_h0} to obtain the $1\sigma$ uncertainty corresponding to Hubble constant expressed in Eqn.~\ref{h_z_0}.

\section{Data}
\label{data}
In our analysis, we use a total $55$ $H(z)$ measurements~\cite{Sharov_2018} estimated by two different techniques, i.e., DA $\&$ BAO. In Table~\ref{table_Hz}, we show these Hubble parameter data in $\rm{km Mpc^{-1}sec^{-1}}$ unit.
\begin{table*}[h]
\caption{\label{table_Hz}Hubble parameter ($H(z)$) measurements and corresponding uncertainties ($\sigma_{H(z)}$), measured by DA and BAO techniques in $\rm{kmMpc^{-1}sec^{-1}}$ unit, are shown at different redshifts ($z$).}
\begin{indented}
\item[]\begin{tabular}{@{}llll|llll}
\br
$z$ & $H(z)$ & $\sigma_{H(z)}$ & Method \& Ref. & $z$ & $H(z)$ & $\sigma_{H(z)}$ & Method \& Ref.\\
\mr
$0.07$ & $69$ & $19.6$ & DA~\cite{Zhang_2014} & $0.48$ &$87.79$ & $2.03$ & BAO~\cite{Wang_2017}\\
$0.09$ & $69$ & $12$ & DA~\cite{Jimenez_2003} & $0.51$ & $90.4$ & $1.9$ & BAO~\cite{Alam_2017}\\
$0.12$ & $68.6$ & $26.2$ & DA~\cite{Zhang_2014} & $0.52$ & $94.35$ & $2.64$ & BAO~\cite{Wang_2017}\\
$0.17$ & $83$ & $8$ & DA~\cite{Simon_2005} & $0.56$ & $93.34$ & $2.3$ & BAO~\cite{Wang_2017}\\
$0.1791$ & $75$ & $4$ & DA~\cite{Moresco_2012} & $0.57$ & $96.8$ & $3.4$ & BAO~\cite{Anderson_2014}\\
$0.1993$ & $75$ & $5$ & DA~\cite{Moresco_2012} & $0.59$ & $98.48$ & $3.18$ & BAO~\cite{Wang_2017}\\
$0.2$ & $72.9$ & $29.6$ & DA~\cite{Zhang_2014} & $0.5929$ & $104$ & $13$ & DA~\cite{Moresco_2012}\\
$0.24$ & $79.69$ & $2.99$ & BAO~\cite{Gazta_aga_2009} & $0.6$ & $87.9$ & $6.1$ & BAO~\cite{Blake_2012}\\
$0.27$ & $77$ & $14$ & DA~\cite{Simon_2005} & $0.61$ & $97.3$ & $2.1$ & BAO~\cite{Alam_2017}\\
$0.28$ & $88.8$ & $36.64$ & DA~\cite{Zhang_2014} & $0.64$ & $98.82$ & $2.98$ & BAO~\cite{Wang_2017}\\
$0.30$ & $81.7$ & $6.22$ & BAO~\cite{Oka_2014} & $0.6797$ & $92$ & $8$ & DA~\cite{Moresco_2012}\\
$0.31$ & $78.18$ & $4.74$ & BAO~\cite{Wang_2017} & $0.73$ & $97.3$ & $7$ & BAO~\cite{Blake_2012}\\
$0.34$ & $83.8$ & $3.66$ & BAO~\cite{Gazta_aga_2009} & $0.7812$ & $105$ & $12$ & DA~\cite{Moresco_2012}\\
$0.35$ & $82.7$ & $8.4$ & BAO~\cite{Chuang_2013} & $0.8754$ & $125$ & $17$ & DA~\cite{Moresco_2012}\\
$0.3519$ & $83$ & $14$ & DA~\cite{Moresco_2012} & $0.88$ & $90$ & $40$ & DA~\cite{Stern_2010}\\
$0.36$ & $79.94$ & $3.38$ & BAO~\cite{Wang_2017} & $0.9$ & $117$ & $23$ & DA~\cite{Simon_2005}\\
$0.38$ & $81.5$ & $1.9$ & BAO~\cite{Alam_2017} & $1.037$ & $154$ & $20$ & DA~\cite{Moresco_2012}\\
$0.3802$ & $83$ & $13.5$ & DA~\cite{Moresco_2016} & $1.3$ & $168$ & $17$ & DA~\cite{Simon_2005}\\
$0.4$ & $95$ & $17$ & DA~\cite{Simon_2005} & $1.363$ & $160$ & $33.6$ & DA~\cite{Moresco_2015}\\
$0.4$ & $82.04$ & $2.03$ & BAO~\cite{Wang_2017} & $1.43$ & $177$ & $18$ & DA~\cite{Simon_2005}\\
$0.4004$ & $77$ & $10.2$ & DA~\cite{Moresco_2016} & $1.53$ & $140$ & $14$ & DA~\cite{Simon_2005}\\
$0.4247$ & $87.1$ & $11.2$ & DA~\cite{Moresco_2016} & $1.75$ & $202$ & $40$ & DA~\cite{Simon_2005}\\
$0.43$ & $86.45$ & $3.97$ & BAO~\cite{Gazta_aga_2009} & $1.965$ & $186.5$ & $50.4$ & DA~\cite{Moresco_2015}\\
$0.44$ & $84.81$ & $1.83$ & BAO~\cite{Wang_2017} & $2.3$ & $224$ & $8$ & BAO~\cite{Busca_2013}\\
$0.4497$ & $92.8$ & $12.9$ & DA~\cite{Moresco_2016} & $2.33$ & $224$ & $8$ & BAO~\cite{Bautista_2017}\\
$0.47$ & $89$ & $34$ & DA~\cite{Ratsimbazafy_2017} & $2.34$ & $222$ & $7$ & BAO~\cite{Delubac_2015}\\
$0.4783$ & $80.9$ & $9$ & DA~\cite{Moresco_2016} & $2.36$ & $226$ & $8$ & BAO~\cite{Font_Ribera_2014}\\
$0.48$ & $97$ & $62$ & DA~\cite{Stern_2010} & & & & \\
\br
\end{tabular}
\end{indented}
\end{table*}

\subsection{DA technique}
\label{da_tech}
One of the methods for measuring $H(z)$ is differential dating of {\it cosmic chronometers} which is suggested by Jimenez \& Loeb (2002)~\cite{Jimenez_2002}. This process utilizes the differential relation between Hubble parameter and redshift, which is given by
\begin{eqnarray}
H(z) & = & -\frac{1}{(1+z)}\frac{dz}{dt}, \label{H_dz}
\end{eqnarray}
where $dz/dt$ is the time derivative of redshift.

{\it Cosmic chronometers} are the selected galaxies showing similar metalicities and low star formation rates. The best {\it cosmic chronometers} are those galaxies which evolve passively on a longer time scale than age differences between them. Age difference ($\Delta t$) between two passively evolving galaxies can be measured by observing $D4000$ break at $4000${\AA} in galaxy spectra~\cite{Moresco_2016}. Measuring the age difference ($\Delta t$) between these types of galaxies, separated by a small redshift interval ($\Delta z$), one can infer the derivative $dz/dt$ from the ratio between $\Delta z$ and $\Delta t$. Then, substituting the value of $dz/dt$ in Eqn.~\ref{H_dz}, we can estimate the Hubble parameter ($H(z)$) at an effective redshift ($z$). This technique measured the values of $H(z)$ in the redshift range $0.07 \leq z \leq 1.965$. Some of these $H(z)$ measurements using this DA technique contain large standard deviations, since the potential systematic errors arise in the measurements due to the contaminations in the selection of cosmic chronometers~\cite{Moresco_2018}.

\subsection{BAO technique}
\label{bao_tech}
The most recent method for measuring $H(z)$ at a particular redshift is the observation of peak in matter correlation function due to baryon acoustic oscillations in the pre-recombination epoch~\cite{Delubac_2015}. Angular separation of BAO peak, at a redshift $z$, is given by $\Delta\theta=r_{d}/(1+z)D_{A}$, where $r_{d}$ is the sound horizon at drag epoch and $D_{A}$ is the angular diameter distance. The redshift separation of BAO peak, at a particular redshift $z$, can be expressed as $\Delta z=r_{d}/D_{H}$, where $D_{H}$ is Hubble distance ($c/H$). Measuring the BAO peak position at any redshift $z$, we can obtain $H(z)$ from the determination of $D_{H}/r_{d}$ and $D_{A}/r_{d}$. BAO $H(z)$ were measured in the redshift range $0.24 \leq z \leq 2.36$. The standard deviations of these $H(z)$ measurements using BAO technique are smaller, since this approach encounters low systematic errors in the measurements of $H(z)$.

\section{Analysis and Results}
\label{results}
%\subsection{Three-point statistics}
\subsection{Sample specific values and uncertainties}
\label{num_val}
\begin{figure*}[h]
%\centering
\includegraphics[scale=0.39]{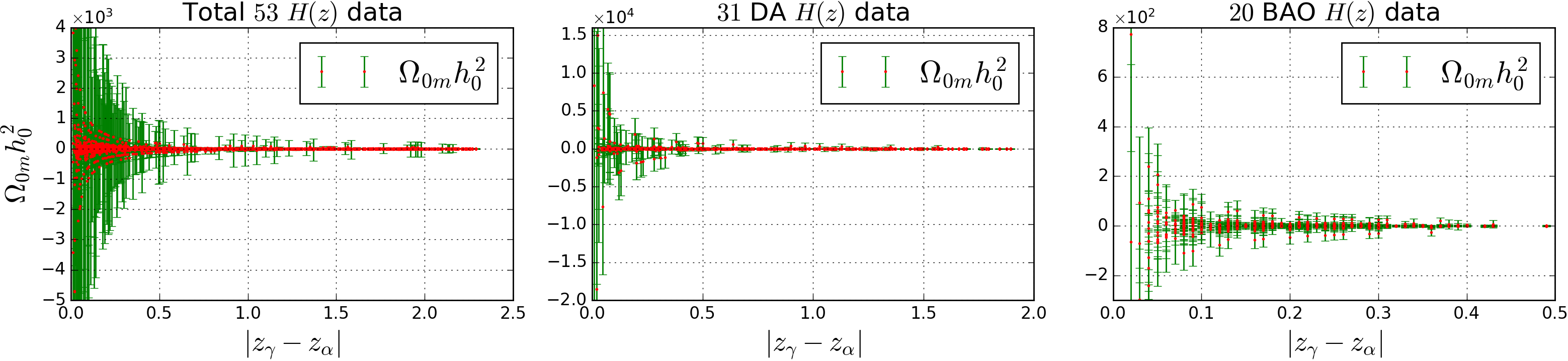}
\caption{Each sub-figure shows the representation of the sample specific values (red) of matter density parameter ($\Omega_{0m}h_{0}^{2}$) with corresponding uncertainties (green) for a particular set of $H(z)$ measurements. Left sub-figure is corresponding to DA+BAO set, middle sub-figure represents the values corresponding to DA set and the right sub-figure shows the distribution of $\Omega_{0m}h_{0}^{2}$ for BAO set. Horizontal axis of each sub-figure defines the absolute values of the differences between two redshifts.}
\label{Omh2_plot}
\end{figure*}
\begin{figure*}[h]
%\centering
\includegraphics[scale=0.39]{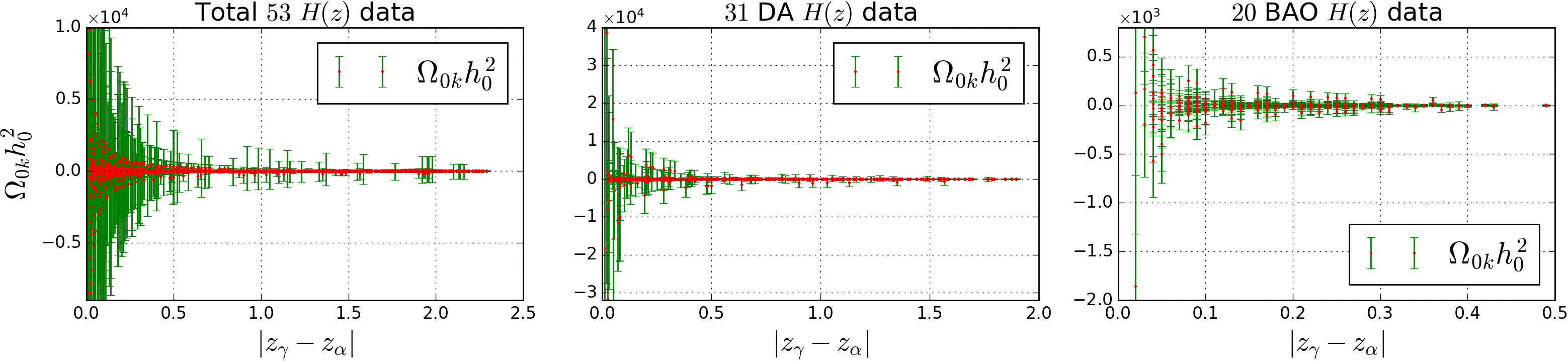}
\caption{Left sub-figure shows the representation of the sample specific values (red) of curvature density parameter ($\Omega_{0k}h_{0}^{2}$) with corresponding uncertainties (green) for DA+BAO set. Similarly, middle and right sub-figures represent the sample specific values (red) of $\Omega_{0k}h_{0}^{2}$ for DA set and BAO set. In each sub-figure, horizontal axis represents the absolute values of the differences between two redshifts.}
\label{Okh2_plot}
\end{figure*}
\begin{figure*}[h]
%\centering
\includegraphics[scale=0.39]{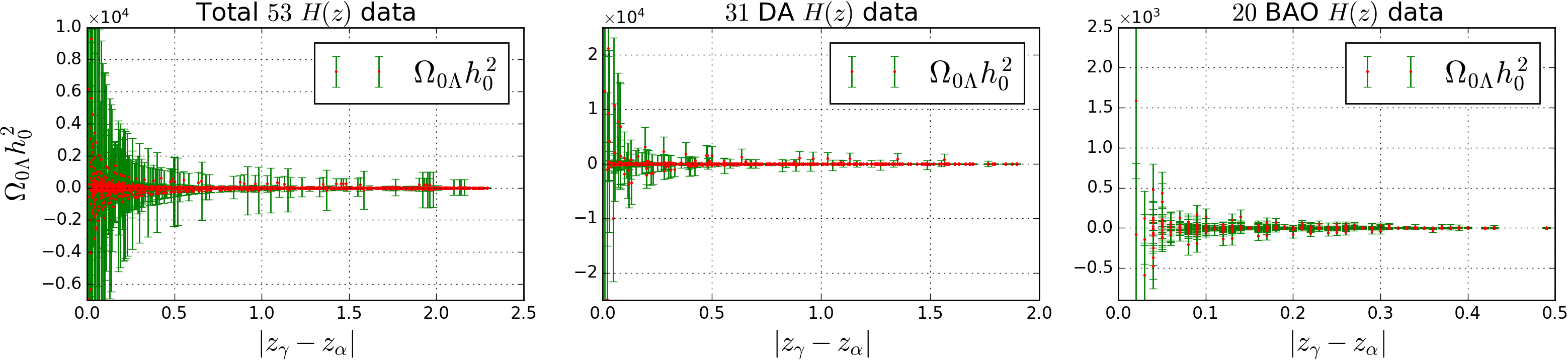}
\caption{Each sub-figure represents the sample specific values (red) of cosmological constant density parameter ($\Omega_{0\Lambda}h_{0}^{2}$) with corresponding uncertainties (green) for a particular set of $H(z)$ measurements. Three sub-figures (left, middle and right) are corresponding to DA+BAO, DA and BAO sets respectively. The absolute values of the differences between two redshifts are presented on the horizontal axis of each sub-figure.}
\label{Odh2_plot}
\end{figure*}
We employ our cosmological parameters estimation procedure (three-point statistics) for three different sets of $H(z)$ measurements. The first data set comprises all Hubble measurements and is denoted as DA+BAO henceforth. The second and third set respectively contains 31 DA measurements (i.e.,DA set) and 20 BAO measurements (i.e., BAO set). The redshift range of DA+BAO Hubble data is $0.07 \leq z \leq 2.36$. DA set contains the redshift range $0.07 \leq z \leq 1.965$. In case of BAO set, we choose the redshift range $0.24 \leq z \leq 0.73$ excluding four highest redshift points, since the non-availability of Hubble data in the large redshift interval between 0.73 and 2.3 affect the best estimates of the cosmological density parameters. If we include these four highest redshift points in the BAO set, we do not get any sample specific values in the interval $0.49 < |z_{\gamma}-z_{\alpha}| < 1.63$ since there is no observed BAO data in a large redshift interval $0.73 < z < 2.3$. In fact, the missing sample specific values would have been more accurate had they been available since they have lower error bars (e.g., please see Figs.~\ref{Omh2_plot},~\ref{Okh2_plot} and~\ref{Odh2_plot} for the generic variation of errors with redshift differences for different cosmological parameters). Due to lack of accurate sample specific values caused by the unavailability of the BAO $H(z)$ measurements for the redshift range $0.73<z<2.3$ we exclude the four large redshift BAO data when using only-BAO set in the analysis. For DA+BAO set each of the two redshifts 0.4 and 0.48 possess two measurements of Hubble parameters (from DA and BAO respectively). For these two redshifts, we include BAO measurements due to their lower measurement errors\footnote{Removing the degenerate redshift points by picking only one value also helps to avoid null values of $A(z_{\alpha},z_{\beta},z_{\gamma})$, e.g., Eqn.~\ref{A_z1z2z3}.}. This results in 53 Hubble measurements in BAO+DA set. We generate the sample specific values (using Eqns.~\ref{Omh2},~\ref{Okh2} and~\ref{Odh2}) of three-point statistics $\Omega_{0m}h_{0}^{2}(z_{\alpha},z_{\beta},z_{\gamma})$, $\Omega_{0k}h_{0}^{2}(z_{\alpha},z_{\beta},z_{\gamma})$ and $\Omega_{0\Lambda}h_{0}^{2}(z_{\alpha},z_{\beta},z_{\gamma})$ as well as the uncertainties (using Eqns.~\ref{sigma_Omh2},~\ref{sigma_Okh2} and~\ref{sigma_Odh2}) corresponding to these sample specific values for each of these three sets\footnote{Here, $z_{\alpha}$, $z_{\beta}$ and $z_{\gamma}$ are three different redshift points (e.g., see section\ref{three_point}).}. We show the sample specific values (with errorbars), with respect to the absolute values of redshift differences $|z_{\gamma}-z_{\alpha}|$, in Figs.~\ref{Omh2_plot},~\ref{Okh2_plot} and~\ref{Odh2_plot} for three cosmological density parameters respectively.
 
We notice a common feature of the sample specific values for the DA set when compared against the same from the full DA+BAO set for each of the three density parameters. The DA set shows relatively larger dispersions of the sample specific values since it contains two DA data at redshift 0.4 and 0.48 with larger error bars than the DA+BAO set. We note in passing that the intrinsic error bars of Hubble measurements from the DA set are larger than the BAO set. Due to the same reason the BAO set shows least dispersions in the sample specific values for all three density parameters when compared with the DA+BAO and DA sets.

We notice from these figures (Figs.~\ref{Omh2_plot}~\ref{Okh2_plot} and~\ref{Odh2_plot}) that the sample specific values show large dispersions for each of the density parameters corresponding to every set of $H(z)$ measurements. The large deviations in the sample specific values of the cosmological parameters arise since the measured Hubble values at different redshifts inevitably contain certain errors. We recall that, from Eqns.~\ref{Omh2},~\ref{Okh2} and~\ref{Odh2} the sample values of the cosmological parameters will be exactly the same as their corresponding true values only for ideal measurements of $H(z)$ free from any error. Even a small error present in the measurements can lead to large values of the sample values of the cosmological parameters specifically when $A(z_{\alpha},z_{\beta},z_{\gamma})$ in the denominator of the above equations are also small, i.e, when each of the $z_{\alpha}$, $z_{\beta}$ and $z_{\gamma}$ are close together. These features are manifested also in Figs.~\ref{Omh2_plot},~\ref{Okh2_plot} and~\ref{Odh2_plot}. Moreover, looking into the vertical error bars of these figures, we can conclude that the uncertainties of the sample specific values are apparently non-Gaussian in nature for each case. We discuss the non-Gaussian nature of sample specific uncertainties (computed using Eqns.~\ref{sigma_Omh2},~\ref{sigma_Okh2} and~\ref{sigma_Odh2}) in the next section.
\begin{figure*}
\centering
\includegraphics[scale=0.52]{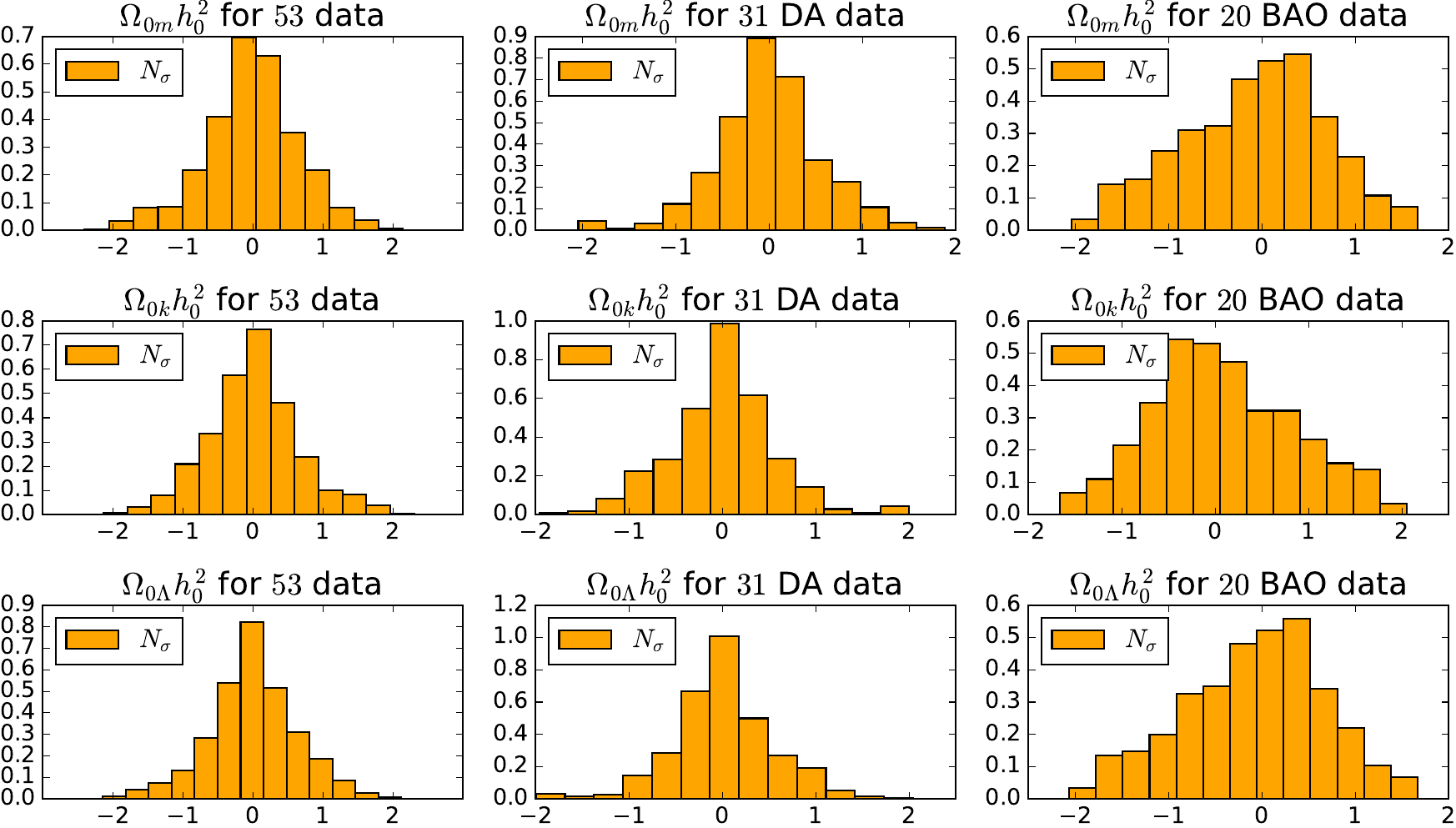}
\caption{Normalized histogram of $N_{\sigma}$ for cosmological density parameters ($\Omega_{0m}h_{0}^{2}$, $\Omega_{0k}h_{0}^{2}$, $\Omega_{0\Lambda}h_{0}^{2}$) corresponding to three sets of $H(z)$ measurements for median statistics. Horizontal axis of each sub-figure represents the dimensionless values of $N_{\sigma}$.}
\label{histogram_md}
\end{figure*}

\subsection{Non-Gaussianity of uncertainties}
\label{non_gauss}
The uncertainties (Eqns.~\ref{sigma_Omh2},~\ref{sigma_Okh2} and~\ref{sigma_Odh2}) corresponding to sample specific values of each density parameter do not possess the Gaussian behaviour. Therefore, we do not use weighted-mean statistics in our analysis, since the sample specific uncertainties are highly non-Gaussian in nature~\cite{Zheng_2016}. However, in case of median statistics, these non-Gaussian uncertainties do not influence the estimated values, since we do not need to use sample specific uncertainties for median analysis. Instead, we directly measure the uncertainties over the estimated cosmological parameters by using Eqn.~\ref{md_var} from sample specific values of these parameters. Chen \etal(2003)~\cite{Chen_2003}, Crandall \& Ratra (2014)~\cite{Crandall_2014} and Crandall \etal(2015)~\cite{Crandall_2015} developed a procedure to measure the non-Gaussianity of sample specific uncertainties. They defined the number of standard deviations ($N_{\sigma}$) which is a measurement of deviation from the central estimation of an observable for a particular statistics. Using the distribution of this number of standard deviations, we can find whether the uncertainties are Gaussian. For instance, the number of standard deviations, corresponding to median ($\rm{md}$) statistics for a particular measurement, is defined as   
\begin{eqnarray}
N_{\sigma, \alpha \beta \gamma} & = & \frac{\Omega_{0x}h_{0}^{2}(z_{\alpha},z_{\beta},z_{\gamma}) - \Omega_{0x}h_{0_{(\rm{md})}}^{2}}{\sigma_{\Omega_{0x}h_{0}^{2}(z_{\alpha},z_{\beta},z_{\gamma})}} \label{N_sigma}
\end{eqnarray}
\begin{table}
%\centering
\caption{\label{N_sigma_md}Percentage of the collection of $N_{\sigma}$ (containing the values within $\pm 1$) for three cosmological density parameters ($\Omega_{0m}h_{0}^{2}$, $\Omega_{0k}h_{0}^{2}$, $\Omega_{0\Lambda}h_{0}^{2}$) corresponding to three sets of $H(z)$ for median (\rm{md}) statistics.}
\begin{indented}
\item[]\begin{tabular}{@{}lccc}
\br
\multicolumn{4}{c}{$\mid N_{\sigma} \mid < 1$ for median (\rm{md}) statistics} \\
\mr
Parameter & DA+BAO set & DA set & BAO set \\
\mr
$\Omega_{0m}h^{2}$ & $86.83 \%$ & $92.52 \%$ & $78.51 \%$ \\
$\Omega_{0k}h^{2}$ & $87.15 \%$ & $92.48 \%$ & $79.03 \%$ \\
$\Omega_{0\Lambda}h^{2}$ & $87.79 \%$ & $93.24 \%$ & $79.91 \%$ \\
\br
\end{tabular}
\end{indented}
\end{table}where `$x$' is anyone of matter ($m$), curvature ($k$) and cosmological constant ($\Lambda$). We use Eqn.~\ref{N_sigma} to calculate the values of $N_{\sigma}$ using sample specific values and our estimated median values for each density parameter for each $H(z)$ set. Then, we collect all $N_{\sigma}$ which satisfy the condition $\mid N_{\sigma} \mid < 1$ for each case. Thereafter, we estimate the percentage of this collection of $N_{\sigma}$ for each density parameter for each $H(z)$ set. If the uncertainty distribution is Gaussian, then the percentage of the collection of $N_{\sigma}$ (containing values within $\pm 1$) should be approximately $68.3 \%$. In Table~\ref{N_sigma_md}, we present our results for the percentage of the collection of $N_{\sigma}$ satisfying the condition $\mid N_{\sigma} \mid < 1$ corresponding to median statistics. We find that percentages of these $N_{\sigma}$ collections for both DA+BAO and DA sets are larger than the same for BAO set. However, the sample specific uncertainties of each density parameter corresponding to each of these three sets do not show the Gaussian nature, since the percentages of $N_\sigma$ collections (containing values within $\pm 1$) for all cases are larger than $68.3\%$ ($1\sigma$ uncertainty). We can also understand the deviation (from Gaussianity) of sample specific uncertainties looking into the probability density of $N_{\sigma}$. In Fig.~\ref{histogram_md}, we show the probability density (normalized histogram) of $N_{\sigma}$ for cosmological density parameters corresponding to three sets of $H(z)$ for median statistics used by us. In this figure, horizontal axis of each sub-figure defines the dimensionless values of $N_{\sigma}$.

\subsection{Density parameters and Hubble constant}
\label{mean_val}
Using median statistics, we estimate the values of three cosmological density parameters ($\Omega_{0m}h_{0}^{2}$, $\Omega_{0k}h_{0}^{2}$, $\Omega_{0\Lambda}h_{0}^{2}$) from the sample specific values of these density parameters and also calculate the uncertainties corresponding to these estimated values using Eqn.~\ref{md_var} for each of the sets (i.e., DA+BAO, DA and BAO) of $H(z)$ measurements. Using Eqn.~\ref{md_cov}, we calculate the covariances between these estimated density parameters for each of three $H(z)$ sets. Thereafter, using the median values, the corresponding uncertainties and covariances of these density parameters, we obtain the value of Hubble constant ($h_{0}$) as well as the corresponding uncertainty for each of three cases by using Eqns.~\ref{h_z_0} and~\ref{sigma_h0} respectively. In Table~\ref{table_cov}, we show the variances (estimated by using Eqn.~\ref{md_var}) corresponding to the median values of density parameters and covariances (estimated by using Eqn.~\ref{md_cov}) between the median values of density parameters for each of three sets of Hubble data.
\begin{table}[h]
%\centering
\caption{\label{table_cov}Table shows the variances of cosmological density parameters as well as the covariances between these density parameters for three different sets of Hubble measurements.}
\begin{indented}
\item[]\begin{tabular}{@{}lccc}
\br
Variance & DA+BAO & DA & BAO \\
\mr
$\sigma^{2}_{\Omega_{0m}h_{0}^{2}}$ & $4.2 \times 10^{-5}$ & $4 \times 10^{-4}$ & $0.0112$\\[2pt]
$\sigma^{2}_{\Omega_{0k}h_{0}^{2}}$ & $2.9 \times 10^{-4}$ & $2.5 \times 10^{-3}$ & $0.0553$\\[2pt]
$\sigma^{2}_{\Omega_{0\Lambda}h_{0}^{2}}$ & $2.1 \times 10^{-4}$ & $1.6 \times 10^{-3}$ & $0.0269$\\
\mr
Covariance & DA+BAO & DA & BAO \\
\mr
${\rm Cov}\bigl(\Omega_{0m}h_{0}^{2},\Omega_{0k}h_{0}^{2}\bigr)$ & $-1.1 \times 10^{-4}$ & $-10^{-3}$ & $-0.0248$\\[2pt]
${\rm Cov}\bigl(\Omega_{0m}h_{0}^{2},\Omega_{0\Lambda}h_{0}^{2}\bigr)$ & $9.3 \times 10^{-5}$ & $8 \times 10^{-4}$ & $0.0177$\\[2pt]
${\rm Cov}\bigl(\Omega_{0k}h_{0}^{2},\Omega_{0\Lambda}h_{0}^{2}\bigr)$ & $-2.5 \times 10^{-4}$ & $-2 \times 10^{-3}$ & $-0.0387$\\
\br
\end{tabular}
\end{indented}
\end{table}
\begin{table}[h]
%\centering
\caption{\label{mean_values}Estimated median values of three cosmological density parameters ($\Omega_{0m}h_{0}^{2}$, $\Omega_{0k}h_{0}^{2}$, $\Omega_{0\Lambda}h_{0}^{2}$) and  Hubble constant ($\hat{h}_{0}$) corresponding to DA+BAO, DA and BAO sets are shown with $68\%$ confidence interval. Last column of the table shows the significances of our estimated parameter values for first two cases and deviations of the same for last case comparing with the Planck results. Detailed descriptions of the results are given in section~\ref{mean_val}.}
\begin{indented}
\item[]\begin{tabular}{@{}lcc}
\br
\multicolumn{3}{c}{DA+BAO set ($0.07 \leq z \leq 2.36$)}\\
\mr
parameter & median value ($1\sigma$ error range) & significance\\
\mr
$\Omega_{0m}h_{0}^{2}$ & $0.1485 \pm 0.0065$ & $0.85\sigma$\\[2pt]
$\Omega_{0k}h_{0}^{2}$ & $-0.0137 \pm 0.017$ & $0.84\sigma$\\[2pt]
$\Omega_{0\Lambda}h_{0}^{2}$ & $0.3126 \pm 0.0145$ & $0.13\sigma$\\[2pt]
$\hat{h}_{0}$ & $0.6689 \pm 0.0021$ & $2.24\sigma$\\
\br
\multicolumn{3}{c}{DA set ($0.07 \leq z \leq 1.965$)}\\
\mr
parameter & median value ($1\sigma$ error range) & significance\\
\mr 
$\Omega_{0m}h_{0}^{2}$ & $0.167 \pm 0.02$ & $1.2\sigma$\\[2pt]
$\Omega_{0k}h_{0}^{2}$ & $0.001 \pm 0.05$ & $0.01\sigma$\\[2pt] 
$\Omega_{0\Lambda}h_{0}^{2}$ & $0.2668 \pm 0.04$ & $1.1\sigma$\\[2pt] 
$\hat{h}_{0}$ & $0.6594 \pm 0.0076$ & $1.9\sigma$\\
\br
\multicolumn{3}{c}{BAO set ($0.24 \leq z \leq 0.73$)}\\
\mr
parameter & median value ($1\sigma$ error range) & deviation\\
\mr 
$\Omega_{0m}h_0^{2}$ & $0.291 \pm 0.106$ & $1.4\sigma$\\[4pt] 
$\Omega_{0k}h_0^{2}$ & $-0.3769 \pm 0.2352$ & $1.6\sigma$\\[4pt]
$\Omega_{0\Lambda}h_0^{2}$ & $0.5269 \pm 0.164$ & $1.3\sigma$\\[4pt] 
$\hat{h}_{0}$ & $0.664 \pm 0.032$ & $0.3\sigma$\\
\br
\end{tabular}
\end{indented}
\end{table}%%%%%%%%%%%%%%%%%%%%%%%%%%%%%%%%%%%%
\begin{table}
%\centering
\caption{\label{cosmo_param}Table shows the cosmological parameters obtained by the Planck Collaboration VI (2020)~\cite{Planck_2018} from standard $\Lambda$CDM model. $\Omega_{0m}$, $\Omega_{0k}$, $\Omega_{0\Lambda}$ are matter, curvature, and cosmological density parameters. $h_0$ is Hubble constant in $100 \ \rm{kmMpc^{-1}sec^{-1}}$ unit.}
\begin{indented}
\item[]\begin{tabular}{@{}lc}
\br
Parameter & Planck's Value \\
\mr
$\Omega_{0m}h_{0}^{2}$ & $0.1430 \pm 0.0011$ \\[2pt]
$\Omega_{0k}$ & $0.001 \pm 0.002$ \\[2pt]
$\Omega_{0\Lambda}$ & $0.6847 \pm 0.0073$ \\[2pt]
$h_0$ & $0.6736 \pm 0.0054$ \\
\br
\multicolumn{2}{c}{Combining the value of $\Omega_{0k}$ and $\Omega_{0\Lambda}$ with the value of $h_0$}\\
\mr
Parameter & Value \\
\mr
$\Omega_{0k}h_{0}^{2}$ & $0.0005 \pm 0.0009$ \\[2pt]
$\Omega_{0\Lambda}h_{0}^{2}$ & $0.3107 \pm 0.006$ \\
\br
\end{tabular}
\end{indented}
\end{table}%%%%%%%%%%%%%%%%%%
In Table~\ref{mean_values}, we present our estimated values of four cosmological parameters ($\Omega_{0m}h_{0}^{2}$, $\Omega_{0k}h_{0}^{2}$, $\Omega_{0\Lambda}h_{0}^{2}$ and $\hat{h}_{0}$) with corresponding uncertainties for three sets of Hubble data. In the same table, we also show the deviation of our estimated values of cosmological parameters from the values of same cosmological parameters constrained by the Planck Collaboration VI (2020)~\cite{Planck_2018}. In Table~\ref{cosmo_param}, we show these parameter values with $1\sigma$ uncertainties obtained by the Planck Collaboration VI (2020)~\cite{Planck_2018}.

\begin{figure*}
\centering
\includegraphics[scale=0.55]{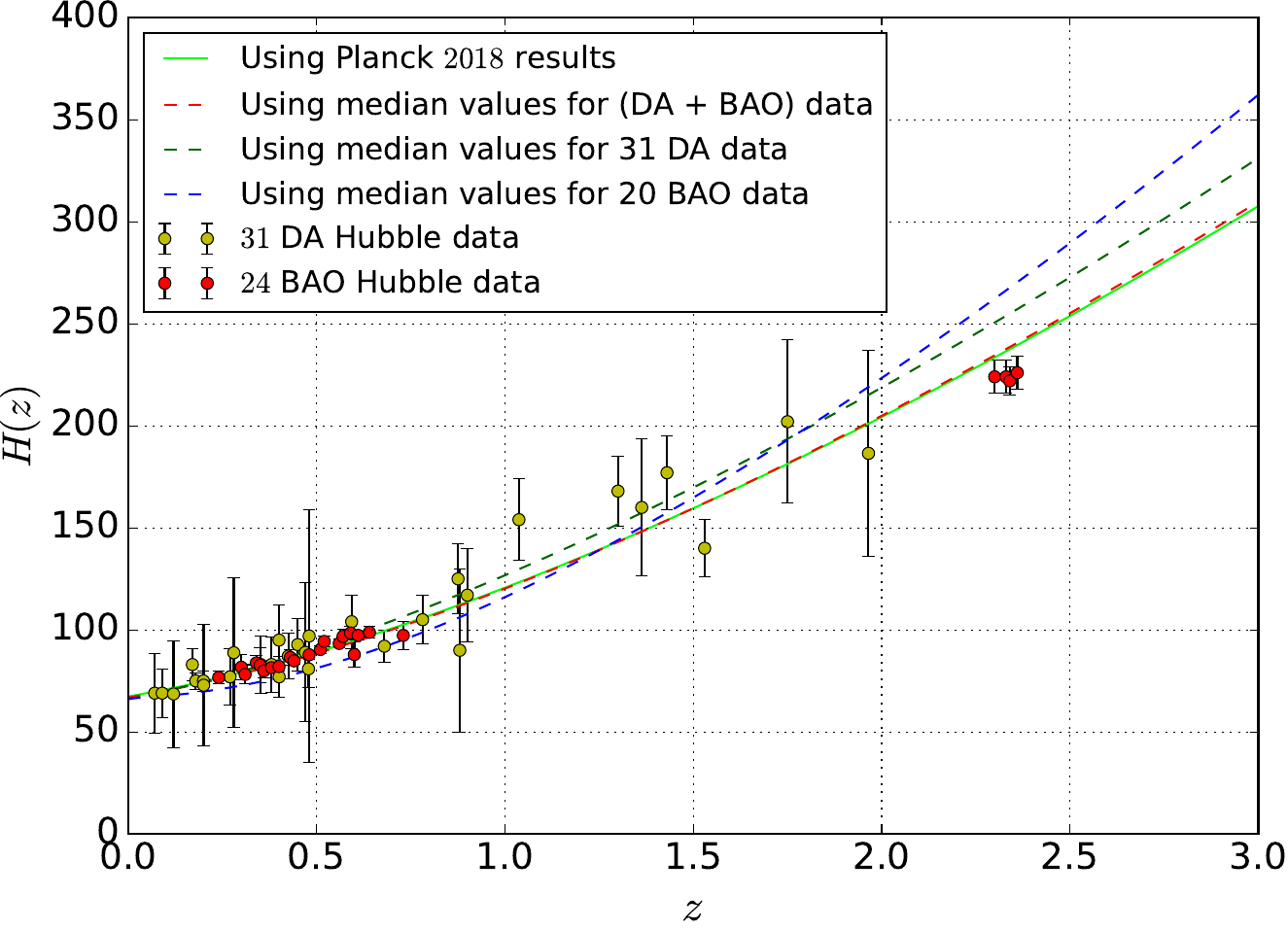}
\caption{Figure shows Hubble parameter ($H(z)$) curves using our estimated values of cosmological parameters as well as using the Planck Collaboration VI (2020)~\cite{Planck_2018} results. The horizontal axis of the figure defines the values of redshift. Moreover, available $55$ $H(z)$ measurements ($31$ DA (yellow) $\&$ $24$ BAO (red)) are also shown in the same figure. Detailed discussions about these $H(z)$ curves are given in section~\ref{mean_val}.}
\label{hubble_plot}
\end{figure*}

For the DA+BAO set, estimated median values of these cosmological parameters are nicely consistent with the values of these four parameters constrained by the Planck Collaboration VI (2020)~\cite{Planck_2018}. The median values of $\Omega_{0m}h_{0}^{2}$ and $\Omega_{0k}h_{0}^{2}$ show $0.85\sigma$ and $0.84\sigma$ deviations respectively from the values of these parameters estimated by the Planck Collaboration VI (2020)~\cite{Planck_2018}, where $\sigma$ is the uncertainty (towards the Planck's parameter value) corresponding to median value of each parameter. Moreover, we get negative median value of $\Omega_{0k}h_{0}^{2}$ for this set of $H(z)$ data. However, comparing the estimated value of $\Omega_{0k}h_{0}^{2}$ with corresponding uncertainty limits, we can conclude that it indicates nearly spatial flatness of the universe. The median value of $\Omega_{0\Lambda}h_{0}^{2}$ shows less deviation (i.e., $0.13\sigma$) from the Planck result. However, our estimated value of $\hat{h}_{0}$ for this DA+BAO set shows $2.24\sigma$ deviation from the Planck's Hubble constant. The reason of this large deviation is that although our estimated value of $\hat{h}_{0}$ is very close to the Planck's estimation, the uncertainty corresponding to our estimated $\hat{h}_{0}$ is very low compared with the difference between our estimated value and the Planck's result of the Hubble constant. The uncertainties corresponding to the median values of three density parameters are larger than the uncertainties of these parameters estimated by the Planck Collaboration VI (2020)~\cite{Planck_2018}, since currently available Hubble parameter measurements are limited in the redshift range $0.07 \leq z \leq 2.36$ and DA measurements contain large errorbars which are shown in Fig.~\ref{hubble_plot}. 

In case of the DA set, our estimated median values of $\Omega_{0m}h_{0}^{2}$ and $\Omega_{0\Lambda}h_{0}^{2}$ show $1.2\sigma$ and $1.1\sigma$ deviations respectively from the Planck's values of these parameters. However, the median value of $\Omega_{0k}h_{0}^{2}$ shows low deviation (i.e., $0.01\sigma$) and $\hat{h}_{0}$ shows large deviation (i.e., $1.9\sigma$) from the values of these parameters obtained by the Planck Collaboration VI (2020)~\cite{Planck_2018}. We note in passing that most of the parameter values (estimated by us) show larger deviation from the Planck results for the DA set compared with our estimated results for the DA+BAO set, since the $31$ data generate a smaller number of sample specific values (which are used to estimate the values of parameters) of density parameters than the same for the $53$ data. Moreover, the internal contaminations (e.g., large standard deviations) in the DA $H(z)$ data also affect the estimated median values of cosmological parameters using the DA data. Due to these reasons, the estimated uncertainty range of each parameter for the DA set is larger than the same for the DA+BAO set. However, the median values of these four parameters using the DA set also show the consistency (within error ranges corresponding to these median values) with the values of these cosmological parameters obtained by the Planck Collaboration VI (2020)~\cite{Planck_2018}.
\begin{figure*}
\centering
\includegraphics[scale=0.5]{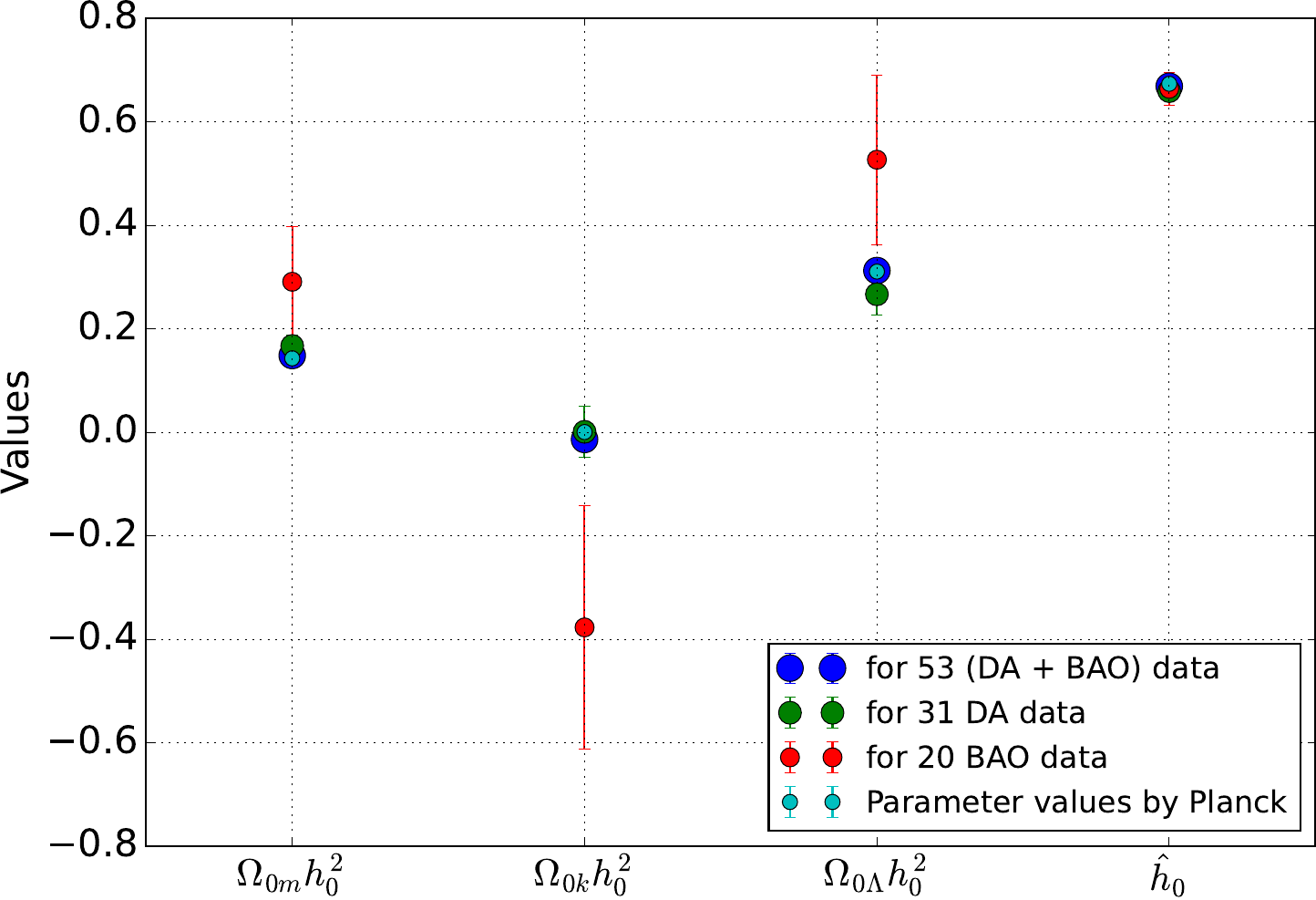}
\caption{Figure shows the estimated median values of cosmological parameters ($\Omega_{0m}h_{0}^{2}$, $\Omega_{0k}h_{0}^{2}$, $\Omega_{0\Lambda}h_{0}^{2}$, $\hat{h}_{0}$) with errorbars. In this figure, we also show the values of same cosmological parameters obtained by the Planck Collaboration VI (2020)~\cite{Planck_2018}. See section~\ref{mean_val} for detailed discussions.}
\label{median_plot}
\end{figure*}

Using $20$ BAO $H(z)$, the estimated median values of most of the cosmological parameters show large deviation (compared with other two sets) from the values of these parameters obtained by the Planck Collaboration VI (2020)~\cite{Planck_2018}. The median values of $\Omega_{0m}h_{0}^{2}$, $\Omega_{0k}h_{0}^{2}$, $\Omega_{0\Lambda}h_{0}^{2}$ and $\hat{h}_{0}$ show $1.4\sigma$, $1.6\sigma$, $1.3\sigma$ and $0.3\sigma$ deviations from the Planck's values of these parameters respectively. Though the BAO technique shows lower standard deviations in $H(z)$ measurements than the DA technique, the available number of $H(z)$ data measured by the former technique is smaller (even lower than the DA $H(z)$ measurements). One reason behind these larger deviations (and large uncertainties of median values) is that the BAO set also produces a smaller number of sample specific values than the same for the DA+BAO set. We note that although the significance of the density parameters for the BAO set is larger than the cases of other two sets, the BAO data alone reduce the deviation of the Hubble constant better than other two sets.

In Fig.~\ref{median_plot}, we show our estimated median values with errorbars of four cosmological parameters ($\Omega_{0m}h_{0}^{2}$, $\Omega_{0k}h_{0}^{2}$, $\Omega_{0\Lambda}h_{0}^{2}$ and $\hat{h}_{0}$) as well as the values (with errorbars) of these parameters constrained by the Planck Collaboration VI (2020)~\cite{Planck_2018}, for better visualisation of the consistency of our estimated values with the Planck results for each of these four cosmological parameters. In this figure, the horizontal axis represents the cosmological parameters and the vertical axis represents the values of these parameters. We note that our estimated median values of these parameters corresponding to each of three Hubble data sets show consistency (within the uncertainty range) with the Planck Collaboration VI (2020)~\cite{Planck_2018} results.

We show the Hubble parameter ($H(z)$) curves in Fig.~\ref{hubble_plot} using the values of cosmological parameters obtained from our analysis for each of three sets of $H(z)$. The horizontal axis of this figure represents the values of redshift. In the same figure, we represent $H(z)$ curve using the density parameter and Hubble constant values constrained by the Planck Collaboration VI (2020)~\cite{Planck_2018}. Moreover, in this figure, we also show the available $31$ DA and $24$ BAO $H(z)$ data points with corresponding uncertainties. the $H(z)$ curve using our median values (corresponding to the DA+BAO set) of cosmological parameters shows excellent agreement with the $H(z)$ curve obtained by using the results of the Planck Collaboration VI (2020)~\cite{Planck_2018}. However, the $H(z)$ curves corresponding to the DA and BAO sets deviate from the Planck $H(z)$ curve, since the estimated parameter values corresponding to the DA and BAO sets contain larger uncertainties. Interestingly, this figure shows that each of three $H(z)$ data sets agree excellently with their corresponding $H(z)$ curve obtained by using corresponding median values of parameters. We also note that, in Fig.~\ref{hubble_plot}, the Hubble curves obtained from our estimated parameter values using DA or BAO data show significant deviations from the Hubble curve using the Planck's cosmological parameters, since the DA $H(z)$ data contain significant internal contaminations and there is non-availability of the BAO data in a large redshift range (i.e., $z<0.24$ and $0.73<z<2.3$). However, utilizing all $H(z)$ data (DA+BAO set) measured by both DA and BAO techniques, our analysis shows an excellent agreement with the Planck Collaboration VI (2020)~\cite{Planck_2018} results constraining spatially flat $\Lambda$CDM model of cosmology.

\section{Discussions and Conclusions}
\label{conclusions}
In this article, we use a new three-point statistics to analyse the available Hubble measurements (DA \& BAO). We use Hubble parameter and redshift relation (Eqn.~\ref{H_z}) and then using the three-point statistics we find the three important cosmological density parameters $\Omega_{0m}h_{0}^{2}$, $\Omega_{0k}h_{0}^{2}$, $\Omega_{0\Lambda}h_{0}^{2}$ in terms of values of redshifts and the corresponding Hubble parameter (e.g., Eqns.~\ref{Omh2},~\ref{Okh2} and~\ref{Odh2}). We also find the uncertainties in the measured values of the density parameters following Eqns.~\ref{sigma_Omh2},~\ref{sigma_Okh2} and~\ref{sigma_Odh2}. We use all possible three-point combinations of the available data points to obtain a set of different values of the density parameters using all or different subsets of the H(z) data. We employ median statistics to estimate the density parameters and the corresponding uncertainties. Using the values of density parameters, we derive the today’s Hubble parameter value and the corresponding uncertainty limits using Eqns.~\ref{h_z_0} and~\ref{sigma_h0}.

Our analysis shows that the uncertainty distributions of the density parameters are highly non-Gaussian. Since a fundamental assumption in using the weighted-mean statistics reliably is the validity of the Gaussian nature of the error distributions, we conclude that it is difficult to interpret these cosmological parameters using weighted-mean statistics in absence of suitable error estimates. \textit{The median statistics on the other hand is appropriate for our analysis, since one does not need to make any assumption about the specific distributions of the errors. Using the median statistics we obtain excellent agreement of the estimates of the density parameters and Hubble constant of this work with those reported by the Planck Collaboration VI (2020)}~\cite{Planck_2018} \textit{(e.g., see Tables~\ref{mean_values} and~\ref{cosmo_param}).} In our analysis, the estimated uncertainties are larger than the uncertainties of parameters calculated by the Planck Collaboration VI (2020)~\cite{Planck_2018}, since the volume of available Hubble data (used by us) is significantly smaller than CMB data.

\textit{It is important to note that although we do not assume a spatially flat universe to begin with, the spatial curvature estimated by us following median statistics using all $53$ Hubble data becomes $-0.0137 \pm 0.017$ indicating a spatially flat universe compatible with the Planck Collaboration VI (2020)}~\cite{Planck_2018} \textit{results. Moreover, the value of today’s Hubble parameter estimated by us $\hat{h}_{0} = 0.6689 \pm 0.0021$ is in close agreement (i.e., $2.24\sigma$ deviation from the Planck's result) with the corresponding value obtained by the Planck Collaboration VI (2020)}~\cite{Planck_2018}. \textit{From Supernovae type Ia measurements, the value of $h_{0}$ is $0.74 \pm 0.014$ (which shows $4.74\sigma$ tension with the Planck's result). Our analysis using Hubble data measured from the relatively local universe therefore is consistent with the global measurement of the same from the CMB data.}

We summarize the major conclusions of our analysis as follows.\\
\hspace*{5pt} (i) We note that the intrinsic errors in Hubble parameter measurements based upon the DA approach are larger than the errors using the BAO measurements. This leads to larger error on estimated cosmological parameters for median statistics using the DA data alone. \textit{However, our estimated parameter values corresponding to DA set are consistent (within our estimated error ranges) with the values of these cosmological parameters constrained by the Planck Collaboration VI (2020)}~\cite{Planck_2018}.\\
\hspace*{5pt} (ii) For the case of 20 BAO data, the uncertainties of cosmological parameters are larger compared with the uncertainties obtained by using other two sets (i.e., 53 DA+BAO and 31 DA data), since the number of $H(z)$ in BAO set for our analysis is lesser than the number of $H(z)$ measurements contained by other two sets. \textit{However, the estimated values of cosmological parameters using BAO set are consistent (within estimated error ranges) with the Planck results.}\\
\hspace*{5pt} (iii) In case of DA+BAO sets, the number of sample specific values of density parameters are larger than the same corresponding to DA and BAO sets, since the large number ($n$) of data generates even larger number ($^{n}C_{3}$) of sample specific values. \textit{Applying median statistics on these sample specific values (for DA+BAO set), we find that the estimated values of fundamental cosmological parameters agree excellently with the Planck Collaboration VI (2020)}~\cite{Planck_2018} \textit{results.}\\
\hspace*{5pt} (iv) \textit{An important advantage of our method is that we estimate the values of fundamental cosmological density parameters without assuming any prior value of any cosmological parameters. Our method provides a unique and yet direct mapping between measured $H(z)$ values and each of the density parameters separately. Thus our method serves as a valuable alternative approach to usual cosmological parameter estimation methods in which all parameters are estimated jointly from the some given data. Such a method is expected to alleviate the problem of degeneracy issues between cosmological parameters during parameter estimation.} We obtain the values of these parameters using our three-point statistics procedure for $\Lambda$CDM model of the universe. In future, we will apply our three-point statistics method in various cosmological models (e.g. wCDM, CPL~\cite{Chevalier_2001,Linder_2003} models) to estimate the fundamental cosmological parameters.

\ack We thank Ujjal Purkayastha, Albin Joseph, Sarvesh Kumar Yadav and Md Ishaque Khan for constructive discussions related to this work.

\section*{Data availability statement}
Data will be shared by authors on receiving request with a specific reason.

\section*{References}
%\begin{harvard}

\end{document}